\documentclass{ws-p8-50x6-00}

\newcommand{\bd}[1]{ \mbox{\boldmath $#1$}  }
\newcommand{\uu}{u({\bf p},s)}
\newcommand{\ubar}{\overline{u}({\bf p'},s')}

\newcommand{\sigvec}{\mbox{\boldmath $\sigma$}}
\newcommand{\etavec}{\mbox{\boldmath $\eta$}}

\newcommand{\kappavec}{\mbox{\boldmath $\kappa$}}

\begin{document}

\title{Relativistic Effects in Electroweak Nuclear Responses}

\author{Maria B. Barbaro}

\address{Dipartimento di Fisica Teorica, Universit\`a di Torino, and INFN,
Sezione di Torino, Via Giuria 1, 10128 Torino, Italy
\\E-mail: barbaro@to.infn.it}

\maketitle

\abstracts{
The electroweak response functions for inclusive electron scattering are
calculated in the Relativistic Fermi Gas model, both in the quasi-elastic and
in the $\Delta$ peak regions. The impact of relativistic
kinematics at high momentum transfer is investigated through an expansion
in the initial nucleonic momentum, which is however exact in the four-momentum
of the  exchanged boson. The same expansion is applied to the meson exchange 
currents in the particle-hole sector: it is shown that the 
non-relativistic currents can be corrected by simple kinematical factors to 
account for relativity. The left-right asymmetry measured via polarized 
electron scattering is finally evaluated in the quasi-elastic and $\Delta$ 
peaks.}

\section{Introduction}
\label{sec:Intro}

The possibility of extracting from polarized electron scattering on protons
the single nucleon strange and axial form factors
has been the focus of many recent studies.
Also the electroexcitation of the $\Delta$ provides important informations on
the very poorly known $N\to\Delta$ transition form factors.
However the scattering on
a free proton is insufficient to disentangle all the form factors
involved in the process: hence the opportunity to perform experiments on 
complex nuclei has been suggested \cite{Bill0}.

Since the double differential asymmetry for inclusive electron scattering

\be
\label{Asym}
{\cal A}=
\frac{d^2\sigma(h=+1)-d^2\sigma(h=-1)}{d^2\sigma(h=+1)+d^2\sigma(h=-1)}
\ee
($h$ being the helicity of the initial electron) is an extremely small 
quantity - roughly of the order of 10$^{-5}$ - 
whose absolute value grows with 
$|Q^2|$, a relativistic treatment of the reaction in both the quasi-elastic 
(QE) and $\Delta$ peak is needed. 

The most appropriate theoretical framework for a consistent study of 
relativistic effects is the Relativistic Fermi Gas (RFG) model, which, in spite
of its simplicity, is capable of well describing the qualitative
features of both peaks, fulfilling the fundamental requirements of {\em 
Lorentz covariance} and {\em gauge invariance}. 
Moreover for a non-interacting system
the RFG yields {\em analytical} expressions for the response functions,
allowing for a clear physical interpretation of each contribution.
On the basis of free RFG nuclear correlations
can then be treated perturbatively, still preserving the two above mentioned 
properties.

\section{Relativistic Fermi Gas response functions}
\label{sec:RFG}

The asymmetry (\ref{Asym}) can be expressed in terms of 5 inclusive response
functions, the longitudinal and transverse electromagnetic $R^L$ and $R^T$
and the longitudinal, 
transverse and axial $\widetilde R^L$, $\widetilde R^T$ and $\widetilde R^{T'}$
parity-violating ones, associated to the interference between the photon
and $Z^0$ exchange, according to

\be
{\cal A} = {\cal A}_0 \frac{v_L \widetilde R^L+
v_T \widetilde R^T + v_{T'} \widetilde R^{T'}}{v_L R^L+v_T R^T} \ ,
\label{AsymRFG}
\ee
$v_{L,T,T'}$ being kinematical factors and 
${\cal A}_0\approx 3.1\times 10^{-4}\tau$.

In the RFG model the nucleus is treated as a collection of free
nucleons described by positive energy Dirac 
spinors with $E(p)=\sqrt{p^2+m_N^2}$: the above responses 
are then analytic and factorize as follows \cite{Bill0,Luis}
\begin{eqnarray}
R^{L,T}(\kappa,\lambda) &=& \frac{3{\cal N}\xi_F}{4\kappa m_N\eta_F^3}
\left[1-\psi^2(\kappa,\lambda)\right] U^{L,T}(\kappa,\lambda) 
\label{respPC}
\\
\widetilde R^{L,T,T'}(\kappa,\lambda) &=&
\frac{3{\cal N}\xi_F}{4\kappa m_N\eta_F^3}
\left[1-\psi^2(\kappa,\lambda)\right] {\widetilde U}^{L,T,T'}(\kappa,\lambda)
\ , 
\label{respPV}
\end{eqnarray}
where ${\cal N}$ is the number
of protons or neutrons, $\kappa=q/2m_N$ and $\lambda=\omega/2m_N$ are
the dimensionless momentum and energy transfers, 
$\eta_F=p_F/m_N$ and $\xi_F=\sqrt{\eta_F^2+1}-1$ the Fermi 
momentum and kinetic energy and

\be
\psi^2(\kappa,\lambda)=\frac{\epsilon_0-1}{\xi_F}=
\frac{1}{\xi_F}\left(\kappa\sqrt{\frac{1}{\tau}+\rho^2}-\lambda\rho-1\right)
\label{psi}
\ee
the squared scaling variable\cite{Alb88,Delta}, linked  to the
minimum energy $\epsilon_0$ required to a nucleon in order to respond to 
the probe. In the above $\tau=\kappa^2-\lambda^2$ and

\be
\rho = 1+\frac{1}{4 \tau}\left(\mu^2-1\right) \ \ \ \ \mbox{with}\ \ \ \ 
\mu=m_\Delta/m_N
\label{rho}
\ee
is an inelasticity parameter which reduces to unity in the 
nucleonic sector, thus allowing for a 
compact treatment of the quasi-elastic and $\Delta$ peaks, providing the 
$\Delta$ is treated as a stable particle: the resonance decay can be 
subsequently taken into account by introducing a finite, energy-dependent, 
width \cite{Delta,Luis,Amo}.

The functions $U$ and $\widetilde U$ in (\ref{respPC},\ref{respPV}) 
depend upon the specific 
electron-nucleon process of interest, although they are not simply
given by single nucleons form factors, since in a relativistic model 
the nucleonic and the many-body content of the problem cannot be factorized.
They read

\begin{eqnarray}
&&U^L(\kappa,\lambda) = 
\frac{\kappa^2}{\tau} \left[(1+\tau\rho^2)
w_2(\tau)-w_1(\tau) +w_2(\tau){\cal D}^L(\kappa,\lambda)
\right] 
\label{UL}\\
&&U^T(\kappa,\lambda) = 2 w_1(\tau) +w_2(\tau){\cal D}^T(\kappa,\lambda) 
\label{UT}\\
&&\widetilde U^L(\kappa,\lambda) = \frac{\kappa^2}{\tau} 
\left[(1+\tau\rho^2) \widetilde w_2(\tau)-\widetilde w_1(\tau) +
\widetilde w_2(\tau){\cal D}^L(\kappa,\lambda)\right] 
\label{ULPV}\\
&&\widetilde U^T(\kappa,\lambda) = 2 \widetilde w_1(\tau) +\widetilde w_2(\tau)
{\cal D}^T (\kappa,\lambda)\label{UTPV}\\
&&\widetilde U^{T'}(\kappa,\lambda) = 2 \sqrt{\tau(\tau\rho^2+1)}
 \widetilde w_3(\tau)\left[1+{\cal D}^{T'}(\kappa,\lambda)\right] 
\label{UTP}\ ,
\end{eqnarray}
where the functions $w_i,{\widetilde w}_i$ contain the single nucleon 
electromagnetic and weak form factors respectively (see 
Refs.~2 and 5 for their expressions). The functions 
${\cal D}^{L,T,T'}$ arise from the relativistic kinematics and read

\begin{eqnarray}
\label{del}
&&{\cal D}^{L,T}(\kappa,\tau) =
\frac{1}{\epsilon_F-\epsilon_0}\int_{\epsilon_0}^{\epsilon_F} 
\eta_T^2(\epsilon)  d\epsilon \\
&&=\frac{\tau}{\kappa^2} \left[ (\lambda\rho+1)^2+(\lambda\rho+1)(1+
\psi^2)\xi_F+\frac{1}{3}(1+\psi^2+\psi^4)\xi_F^2\right] -(1+\tau\rho^2)
\nonumber\\
&&{\cal D}^{T'}(\kappa,\tau) = \frac{1}{\epsilon_F-\epsilon_0}
\int_{\epsilon_0}^{\epsilon_F}
 \left(\sqrt{1+\frac{\eta_T^2(\epsilon)}{1+\tau\rho^2}}-1\right) d\epsilon
\nonumber\\
&&=\frac{1}{\kappa} \sqrt{\frac{\tau}{1+\tau\rho^2}}
\left[1+\xi_F(1+\psi^2)+\lambda\rho\right]-1
\ .
\label{dela}
\end{eqnarray}
The above expressions show that ${\cal D}^L$ and ${\cal D}^T$
just correspond to the mean square value of the transverse momentum 
\be
\eta_T(\epsilon) = 
\sqrt{\frac{\tau}{\kappa^2}(\epsilon+\lambda\rho)^2-1-\tau\rho^2}
\ee
of the nucleon, whereas ${\cal D}^{T'}$ is related
to the transverse kinetic energy of the nucleon when $\tau$ is small.
It is worth noticing that the ${\cal D}$ functions are quadratic in the 
Fermi momentum and vanish in non-relativistic models.

The relativistic response functions thus obtained are sensibly different
from the non-relativistic ones for momentum transfer higher than about 1
GeV/c, the differcence obviously growing with $q$.

As an illustration, we focuss on the $\Delta$ region, where
the transverse Fermi motion gives rise to a longitudinal response function, 
even for a spherical (M1) $\Delta$, as predicted by the constituent quark 
model. This effect, first pointed out in Ref.~6,
is negligible at low $q$ and $k_F$, but becomes significant in heavy nuclei
and for high momentum transfer. Moreover, if one considers the full 
$N\to\Delta$ vertex \cite{Jones}, that includes M1, E2 and C2 amplitudes,
the longitudinal response in the $\Delta$ region turns out to be

\be
R^L_\Delta(\kappa,\lambda)\propto G_{C,\Delta}^2(\tau)+
\frac{4\mu^2\left[G_{M,\Delta}^2(\tau)+3 G_{E,\Delta}^2(\tau)\right]+
G_{C,\Delta}^2(\tau)}{\tau(1+\tau\rho^2)} {\cal D}^L(\kappa,\tau)\ ,
\ee
thus arising from both the Coulomb transition form factor and 
from the many-body relativistic effect contained in ${\cal D}$. This effect,
evaluated in Ref.~\cite{Delta} in the RFG model, yields a $\Delta$ longitudinal
response which is about 15\% of the corresponding quasi-elastic one at 
q=1 GeV/c in absence of Coulomb form factor. 
The latter is shown to further enhance this ratio 
of almost a factor 2, thus showing the strong sensitivity of this response to 
the deformation of the $\Delta$ resonance.

To go beyond the free Fermi gas, the impact of the N-N 
interaction on the response functions should be calculated in the same
relativistic framework. In Ref.\cite{Rinaldo} the effect of nuclear 
correlations on $R^L_\Delta$ is included in the Boson Loop Expansion
framework, which sums up an infinite class of particle-hole, $\Delta$-hole, 
2p-2h and 3p-3h diagrams: the results indicate a substantial
enhancement of the response at $q\simeq$ 500 MeV/c. Unfortunately the 
calculation could not be applied at higher momentum transfers, since it is 
non-relativistic.
Although exact relativistic calculations can in principle be performed, we
illustrate in the next Section a new non-relativistic expansion of the 
nuclear currents, first suggested in Ref.~9, 
which provides an extremely
good approximation to the exact result {\em for any value of the 
energy and momentum transfers} and allows for the inclusion of relativistic
effects in non-relativistic calculations through simple kinematical factors.

\section{A new non-relativistic expansion}
\label{sec:exp}

The method treats asymmetrically the nucleons in initial and final states, 
expanding the single nucleon current in powers of the parameter
$\eta=p/m_N$, where $p$ is 
the three--momentum of the nucleon inside the Fermi sphere, and truncating
the expansion at linear order. Since $\eta\leq \eta_F$, $\eta_F$ being 
tipycally about 1/4, this turns out to be an excellent approximation, which,
compared to the traditional non-relativistic expansions, has the advantage of
retaining the full dependence on $\kappa$ and $\lambda$. The same
procedure can also be applied to 2-body currents, like meson-exchange 
\cite{MEC} and correlation currents \cite{Corr}.
The ``relativized'' currents so obtained can be implemented together with 
relativistic kinematics in standard non-relativistic models
of one-particle emission near the quasi-elastic and $\Delta$ peaks. 

To illustrate the method, let us discuss in some details the single-nucleon 
on-shell electromagnetic current operator 

\be
J^{\mu}(P's';Ps) = 
\ubar \left[ F_1\gamma^\mu + \frac{i}{2m_N}F_2\sigma^{\mu\nu}Q_\nu 
\right] \uu \ .
\label{eq1}
\ee
By expanding the Dirac spinors in powers
of the bound nucleon momentum ${\etavec}={\bf p}/m_N$ one gets (omitting
the spinors $\chi_s$, $\chi^\dagger_{s'}$)

\begin{eqnarray}
J^0 &\simeq& 
	\frac{\kappa}{\sqrt{\tau}}G_E
	+\frac{i}{\sqrt{1+\tau}}\left(G_M-\frac{G_E}{2}\right)
	(\kappavec\times\etavec) 
	\cdot\sigvec
\\
{\bd J}^\perp &\simeq&
\frac{1}{\sqrt{1+\tau}}\left\{
	iG_M(\sigvec\times\kappavec)+
	\left(G_E+\frac{\tau}{2}G_M\right)
	\left(\etavec
	-\frac{\kappavec\cdot\etavec}{\kappa^2}\kappavec\right) 
\right. \nonumber \\
&-& \left.
	\frac{iG_M}{1+\tau}
	(\sigvec\times\kappavec)\kappavec\cdot\etavec 
+	\frac{iG_M}{2(1+\tau)}(\etavec\times\kappavec)
	\sigvec\cdot\kappavec \right\}
\end{eqnarray}
for the time and space transverse components, respectively, the longitudinal 
current being trivially linked to $J^0$ via current conservation. 
By comparing these expressions with the traditional non-relativistic ones

\begin{eqnarray}
{J^0}_{nonrel} &=&  G_E \\
{\bd J}_{nonrel}^\perp &=&
 iG_M(\sigvec\times\kappavec)+
	G_E\left(\etavec-\frac{\kappavec\cdot\etavec}{\kappa^2}
	\kappavec\right)
\label{eq30}
\end{eqnarray}
one notices that, besides the extra terms linear in $\eta$, the
kinematical factor $\kappa/\sqrt{\tau}> 1$ is present in the time component,
corresponding to an {\em enhancement} of the charge response function, whereas 
the factor $1/(1+\tau)< 1$ in ${\bd J}^\perp$ induces a {\em reduction} of the
transverse response. These effects are directly linked to the relativistic
phenomena of length contraction and time dilation.

Although technically more cumbersome because of the spin structure of the 
$\Delta$, the same procedure can be applied to the $N\to\Delta$ transition 
current. If, for simplicity, we restrict our analysis to M1
transistions, the $\eta$-expansion of the current \cite{Jones}

\begin{equation}
J^\mu_{N\to\Delta} (P_\Delta s_\Delta;P s)= 
\overline{u}_\Delta^\beta(P_\Delta,s_\Delta) \frac{G}{2m_N^2} 
\epsilon^{\beta\mu\alpha\gamma} (P_\alpha+P_{\Delta\alpha}) Q_\gamma u(P,s) \ ,
\label{eq3}
\end{equation}
where $u_\Delta^\beta(P_\Delta,s_\Delta)$ is
the Rarita-Schwinger spinor describing a spin-3/2 particle and $G$ is 
proportional to the magnetic transition form factor $G_{M,\Delta}$, yields

\begin{eqnarray}
J^0_{N\to\Delta} &=&
-2 G\sqrt{1+\tau'}{\bf S}^{\dagger}\cdot(\etavec\times\kappavec) +O(\eta^2)
\label{J0-nr}\\
{\bd J}_{N\to\Delta}
&=& 2 G \sqrt{1+\tau'}
    ({\bf S}^{\dagger}\times\kappavec) + O(\eta) \ ,
\label{J-final}
\end{eqnarray}
where $\tau'=[4\tau+(\mu-1)^2]/(4\mu)$ and ${\bf S}$ is the usual 1/2$\to$3/2
spin transition operator.

The quality of the new approximation to the relativistic, magnetic
$\Delta$ current is shown in fig.~1, where the exact RFG longitudinal
and transverse responses obtained using magnetic and electric form factors 
are displayed with solid lines. 
In addition we show with dashed lines the responses computed in
the non-relativistic Fermi gas model with relativistic kinematics
and the new currents in Eqs.~(\ref{J0-nr},\ref{J-final}).
For comparison we also show with dot-dashed lines results for
the non-relativistic Fermi gas model and relativistic kinematics, 
but using the 
traditional non-relativistic current. The improvement of the
description of the relativistic results using our currents is clear
from this figure --- the solid and dashed lines almost coincide.
This proves that our expansion to order $O(\eta)$ is precise enough 
to describe the $\Delta$ excitation in nuclei with negligible error 
for high momentum transfers. 

\begin{figure}
\begin{center}
\includegraphics[width=12cm,
bbllx=70bp,bblly=350,bburx=560,bbury=505,clip=]{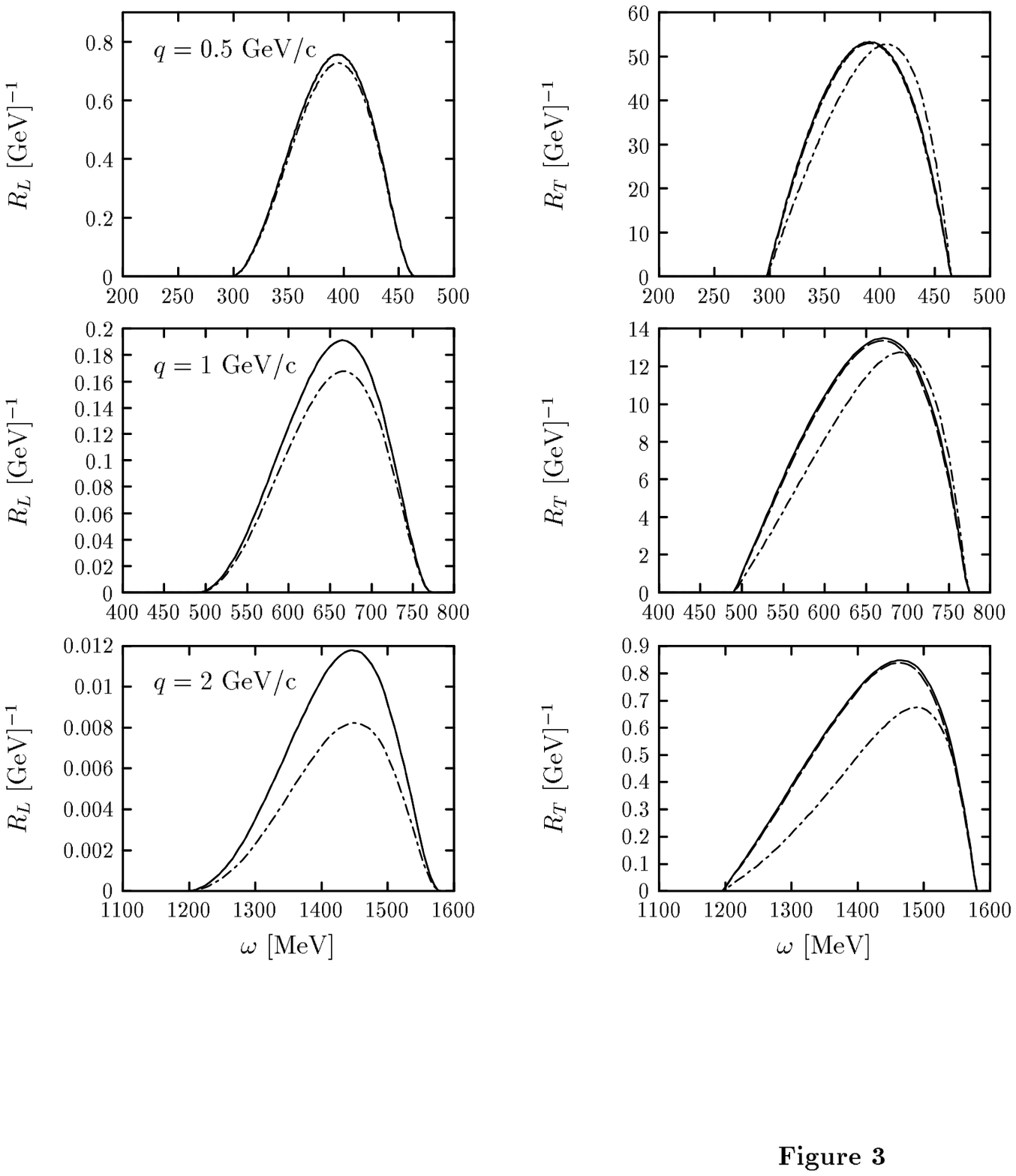}
\caption{Electromagnetic responses in the $\Delta$-peak at $q=$2 GeV/c
for $^{12}$C with $k_F=225$ MeV/c, without a $\Delta$-width.
Solid: exact results within the RFG model;
dashed: new expansion of the electromagnetic current to 
first  order in $\eta$;
dot-dashed: traditional non-relativistic current. 
We use relativistic kinematics in all cases.
}       
\label{figup}
\end{center}
\end{figure}

\section{Meson Exchange Currents}
\label{sec:MEC}

We now consider the 
Meson-Exchange-Currents (MEC) associated to pion exchange, which have to be 
taken into account in order to preserve current conservation. We have
applied the above non-relativistic expansion to 
the particle-hole matrix elements of the so-called pion-in-flight and 
seagull currents,
corresponding to $\gamma\pi\pi$ and $\gamma\pi N$ vertices respectively,
and compared the results with the exact relativistic calculation and with the 
``traditional'' non-relativistic one, where $q$ and $\omega$ are not treated 
exactly \cite{ADM}.

In Fig.~2 we show, for sake of illustration, the p-h matrix element of a
typical component of the pion-in-flight current at $q$=2 GeV/c
in the allowed spin channels (spin diagonal in the left panel and 
spin-flip in the right panel); 
similar results hold for the other components and for the seagull current,
as shown in Ref.~10. Again our 
approximation is in fact indistinguishable from the exact calculation.

Notably the final expressions 
for the ``relativized'' currents (R) is linked to the traditional 
non-relativistic (TNR) ones through simple kinematical factors. For the 
pion-in-flight this is a global reduction factor

\be
J^{\mu,R}_P = \frac{1}{\sqrt{1+\tau}} J_P^{\mu,TNR} 
\ee
and the same relation holds for the time component of the seagull current, 
whereas for its space component we get

\be
{\bd J}^{R}_S = \frac{1}{\sqrt{1+\tau}} {\bd J}_S^{1,TNR}+ 
\sqrt{1+\tau} {\bd J}_S^{2,TNR} \ ,
\ee
where ${\bd J}^1_S$ and ${\bd J}^2_S$ correspond to two different pionic 
momentum flows (see Ref.~10 for details).

\begin{figure}
\begin{center}
\includegraphics[width=12cm,bbllx=70bp,bblly=155,bburx=550,bbury=315,
clip=]{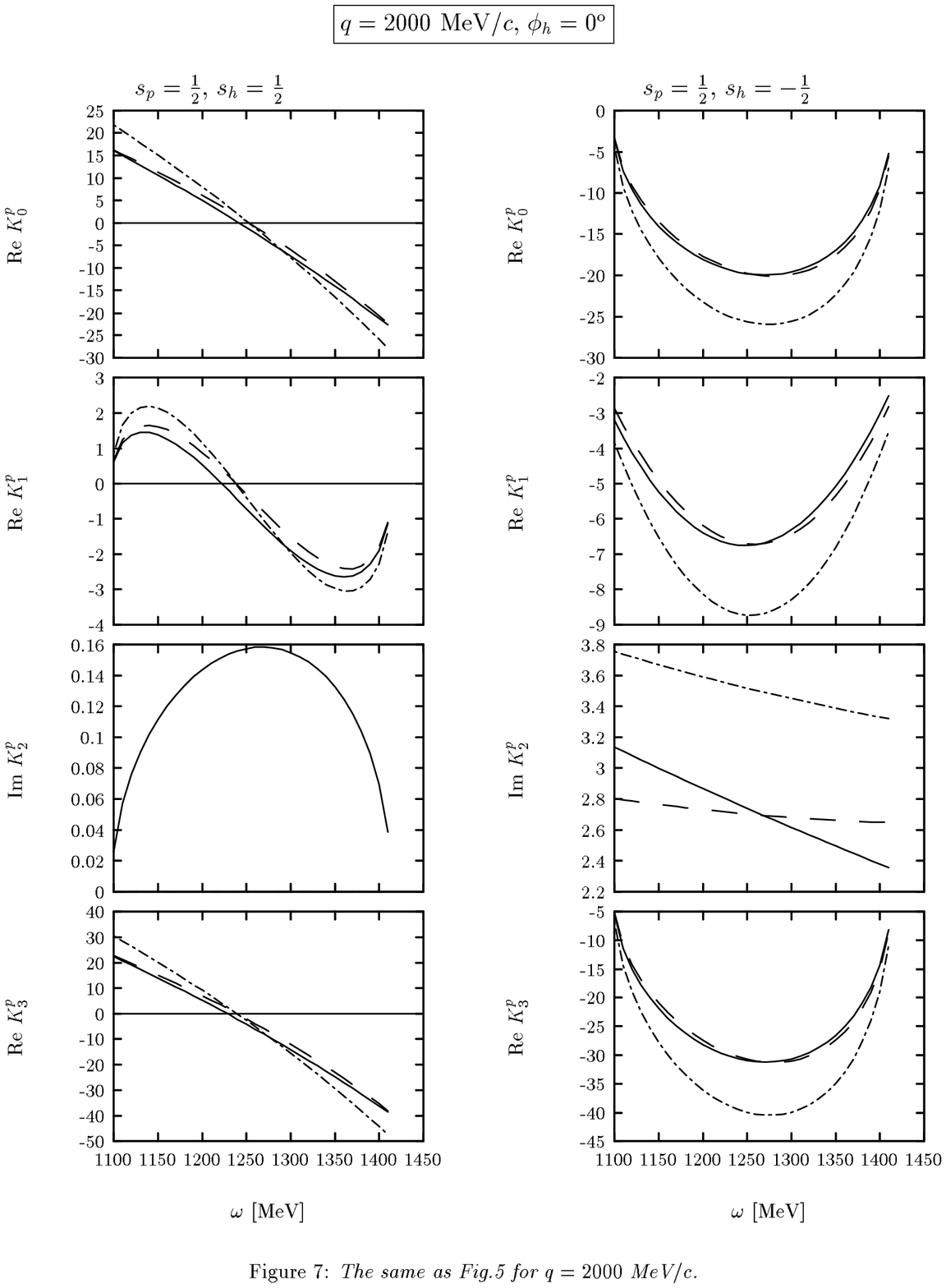}
\caption{Pion-in-flight p-h matrix elemnts (in arbitrary units) for q=2 GeV/5.
The kinematics for the hole is $|\vec h|$=175 MeV/c and the azimuthal angle
is $\Phi=0$. The spins are $s_p=s_h=1/2$ in the left panel and  $s_p=-s_h=1/2$
in the right panel. Soild: relativistic; dashed: our approximation; dot-dashed:
traditional non-relativistic}       
\end{center}
\end{figure}

\section{The asymmetry in the $\Delta$-resonance region}
\label{sec:asym}

Due to the purely {\em isovector} nature of the $\Delta$ resonance, the
asymmetry (\ref{AsymRFG}) reads in this region

\be
{\cal A}_{N-\Delta} = {\cal A}_0 \left\{ - \left(1-2 \sin^2\theta_w\right) + 
v_{T'} \ \frac{\widetilde{R}_{VA}^{T'}(\kappa,\lambda)}{v_L R^L(\kappa,\lambda) + 
v_T R^T(\kappa,\lambda)} \right\} \ ,
\label{eq:asymnd}
\ee
$\theta_w$ being the Weinberg angle.
The above formula clearly shows that if the axial $N \rightarrow \Delta$ 
response can be neglected then the inelastic asymmetry, normalized to 
${\cal A}_0$ and displayed versus $\lambda$ for fixed $\kappa$, would be
flat in the $\Delta$ domain. Hence a departure from flatness would signal 
the presence of the axial response.
In Fig.~3 the ratio ${\cal A}/{\cal A}_0$ is displayed versus
$\omega$ at  $q$=350 MeV/c and $\theta = 10^0$, which are, according 
to the analysis of Ref.~5, the most favourable kinematical conditions: 
indeed the asymmetry is increased by about $10\%$.

\begin{figure}
\begin{center}
\includegraphics[width=10cm,height=6cm]{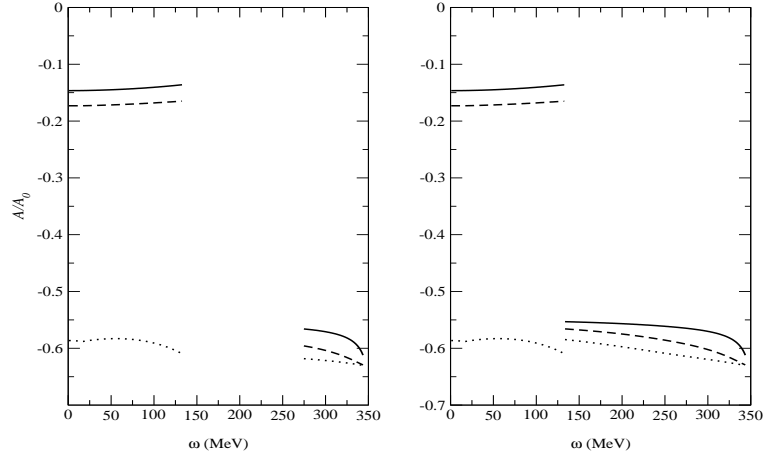}
\caption{Asymmetry at $\theta=10^0$ (solid), $\theta=30^0$ (dashed) and
$\theta=150^0$ (dotted) for $q=350 \ {\rm MeV/c}$. The $\omega$ range encompasses both
the QEP and the $\Delta$ domain. The left and right panels respectively refer 
to a vanishing width and to  a finite decay width $\Gamma(s)$.}       
\end{center}
\end{figure}

The most notable feature of the ${\cal A}$  found relates to the
dramatic increase of its magnitude  as one makes a transition from the QEP 
into the $N \rightarrow \Delta$ region for small electron scattering angles.
Because of its size this effect should be measurable both at large
($\sim$1 GeV/c) and moderate ($\sim$300-400 MeV/c) momentum transfer.
In the former case nuclear interactions are not likely to disrupt the RFG
predictions too much.
In the latter a modification of the effect could take place, but 
then this might eventually help to shed light on the nature of the 
$NN$ and $N\Delta$ forces.

\section*{Acknowledgments}
The work presented is the result of collaborations with 
L. Alvarez-Ruso, J.E. Amaro, P. Amore, J.A. Caballero, R. Cenni, 
T.W. Donnelly and A. Molinari.

\end{document}